\documentclass[runningheads]{llncs}

\usepackage{booktabs}
\usepackage{cite}
\usepackage{xcolor}
\usepackage{amsmath}
\usepackage{amssymb}
\usepackage{tabularx}
\usepackage{array}
\usepackage{cleveref}
\usepackage{multirow}
\usepackage[T1]{fontenc}
\usepackage{graphicx,verbatim}

\begin{document}
\title{PINNOCHIO: Physics-Informed Neural Network for Coupled Hyperelastic Interface-Volume Simulation in Orthognathic Surgery\thanks{This work has been submitted to MICCAI 2026.}}

\titlerunning{PINNOCHIO}

\author{
Jungwook~Lee\inst{1} \and
Daeseung~Kim\inst{2} \and
Kevin~Gu\inst{2} \and
Zhangfeng~Hu\inst{1} \and
Tianshu~Kuang\inst{2} \and
Finn~Hopeman\inst{1} \and
Michael~A.K.~Liebschner\inst{3} \and
Jaime~Gateno\inst{2} \and
Pingkun~Yan\inst{1}\textsuperscript{($\boxtimes$)}
}

\authorrunning{J. Lee et al.}

\institute{
Department of Biomedical Engineering and Center for Biotechnology and Interdisciplinary Studies, Rensselaer Polytechnic Institute, Troy, NY 12180, USA \\
\email{yanp2@rpi.edu}
\and
Department of Oral and Maxillofacial Surgery, Houston Methodist Research Institute, Houston, TX 77030, USA
\and
Department of Neurosurgery, Baylor College of Medicine, Houston, TX 77030, USA
}

\maketitle
\begin{abstract}
Predicting patient-specific facial soft-tissue deformation is critical for iterative orthognathic surgery planning. However, current computational methods face a strict accuracy-efficiency trade-off: high-fidelity Finite Element Methods (FEM) are computationally prohibitive, whereas pure deep learning models often produce biomechanically inconsistent results. While Physics-Informed Neural Networks (PINNs) offer a promising avenue, learning the complex heterogeneous mechanics of bone--soft-tissue interactions with only partial clinical supervision (i.e., outer facial surfaces) remains highly unstable. To overcome these challenges, we present PINNOCHIO, a novel physics-informed framework for facial soft-tissue simulation. PINNOCHIO introduces a hybrid sequential decomposition that explicitly decouples discontinuous bone--soft-tissue interface movements from continuous volumetric hyperelastic deformation. This structural separation enables stable training and facilitates a physics-enabled sim-to-real adaptation strategy, ensuring internal biomechanical consistency without requiring volumetric ground truth. Evaluated on a 40-patient clinical cohort, PINNOCHIO outperforms existing baselines in both surface accuracy and physical validity. Furthermore, it achieves a substantial speedup over FEM, successfully resolving the accuracy-efficiency trade-off to provide a highly reliable and practical tool for interactive surgical planning.
\end{abstract}
\section{Introduction}
Orthognathic surgery aims to correct dentofacial deformities by repositioning the jawbones to restore facial harmony. To achieve optimal outcomes, surgeons must iteratively evaluate multiple candidate plans by simulating the corresponding soft-tissue responses~\cite{dall2021orthognathic,proffit2018contemporary, naini2025facial}. However, this task is fundamentally challenging due to the highly non-linear mechanical relationship between the rigid bone and the surrounding soft tissues~\cite{lampen2022deep,flynn2013simulating}. Consequently, there is a pressing clinical need for simulation tools that balance biomechanical rigor with computational efficiency, enabling interactive planning while reducing manual trial-and-error ~\cite{ruggiero2023soft,lampen2022deep,lampen2023spatiotemporal,alcaniz2021soft}. Current computational methods face a fundamental trade-off between accuracy and efficiency~\cite{alcaniz2021soft}. High-fidelity patient-specific finite element methods (FEM) provide biomechanical rigor but are computationally prohibitive and require extensive subject-specific parameter tuning, limiting their practicality for iterative planning~\cite{otamendi2011designing,lampen2022deep,alcaniz2021soft}. Conversely, pure data-driven deep learning models~\cite{berends2025soft,fang2024correspondence} offer rapid inference speeds but remain physics-agnostic. Lacking physical constraints, these approaches may produce geometrically plausible surfaces, yet they often yield biomechanically inconsistent internal behaviors, such as volumetric distortion or element inversion~\cite{pal2022towards,hu2019preventing}.
\begin{figure}[t]
    \centering
    \includegraphics[width=0.8\textwidth, keepaspectratio]{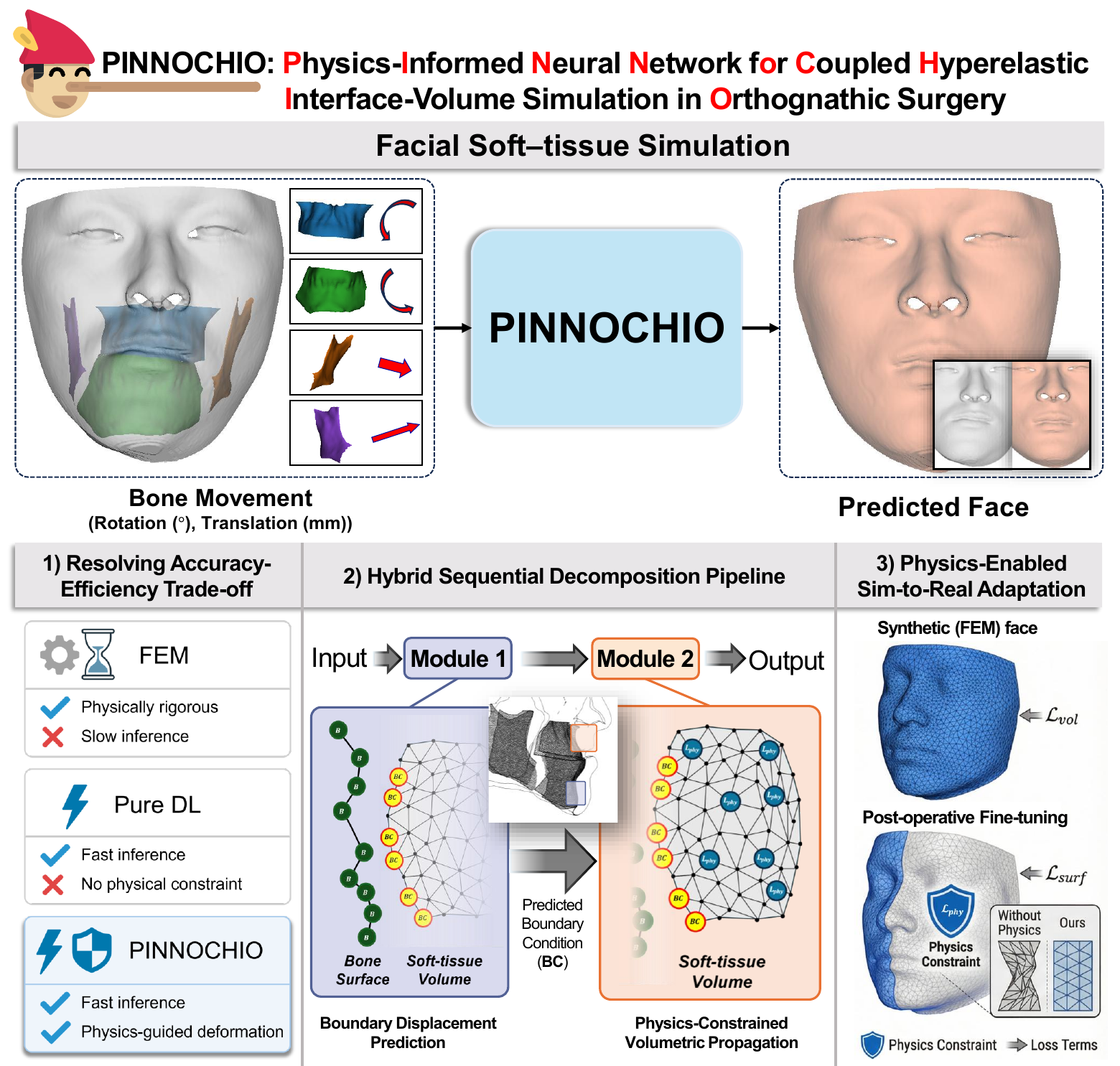}
    \caption{
    PINNOCHIO for facial soft-tissue simulation. \textbf{1)} Accuracy-efficiency trade-off in existing FEM and physics-agnostic deep learning. \textbf{2)} Proposed sequential decomposition: interface prediction followed by physics-constrained volumetric propagation. \textbf{3)} Physics-guided sim-to-real adaptation using surface-only clinical supervision.
    }
    \label{fig:PINNOCHIO_Main}
\end{figure}
Physics-Informed Neural Networks (PINNs)~\cite{raissi2019physics,karniadakis2021physics,raissi2017physics} offer a promising avenue to reconcile this trade-off by incorporating governing physical equations directly into the loss function. However, applying a unified PINN framework to real-world orthognathic planning is hindered by two fundamental barriers. First, surgical deformation exhibits heterogeneous mechanics: discontinuous sliding at the bone--soft-tissue interface and large, continuous hyperelastic deformation within the tissue volume~\cite{kim2021novel}. Jointly optimizing these mechanically distinct regimes within a single PINN often results in unstable training dynamics and degraded convergence~\cite{krishnapriyan2021characterizing}. Second, clinical supervision is inherently limited; while post-operative volumetric imaging (e.g., CT/CBCT) is available, internal tissue motion lacks dense correspondence~\cite{mollemans2007predicting}. This confines supervision primarily to the outer facial surface\cite{awad2022accuracy,franco2025three}, where purely data-driven learning cannot guarantee physically consistent internal fields. Together, these challenges motivate a structured, sequential architecture that (i) predicts interface displacements and (ii) propagates them through the volume with physical constraints, allowing for end-to-end fine-tuning under surface-only clinical supervision. 

To address these challenges, we propose \textbf{PINNOCHIO} (Physics-Informed Neural Network for Coupled Hyperelastic Interface-Volume Simulation in Orthognathic Surgery), a framework for simulating soft-tissue deformation driven by skeletal repositioning that integrates data-driven efficiency with physics-based regularization. Our main contributions are summarized as follows:

\begin{itemize}
    \item[$\bullet$] \textbf{Resolving Accuracy-Efficiency Trade-off}: We present a fast, physics-consistent simulator that bridges the accuracy gap to high-fidelity FEM simulations, enabling rapid and biomechanically valid iterative planning.
    \item[$\bullet$] \textbf{Hybrid Sequential Decomposition}: We propose a sequential decomposition to handle heterogeneous surgical mechanics by explicitly decoupling discontinuous interface motion from continuous volumetric hyperelastic deformation, improving training stability and convergence on complex anatomies.
    \item[$\bullet$] \textbf{Physics-Enabled Sim-to-Real Adaptation}: We introduce an adaptation strategy that bridges synthetic FEM supervision and clinical data by fine-tuning with postoperative outer facial surface supervision only, encouraging internal physical consistency without requiring volumetric ground truth.
\end{itemize}
Quantitative validation on a 40-patient clinical cohort demonstrates that PINNOCHIO outperforms representative baselines in both surface accuracy and physical validity, supporting its clinical viability for real-world surgical planning.
\section{Problem Setup} \label{sec:Problem Setup}
Let $\Omega_0 \subset \mathbb{R}^3$ denote the patient’s preoperative soft-tissue volume (muscle, fat, and skin). 
Our goal is to estimate a 3D displacement field $u$ mapping $\Omega_0$ to its postoperative configuration (\textbf{\Cref{fig:PINNOCHIO_Main}}).
We focus on two clinically meaningful boundary subsets $\partial \Omega_0$: 
(i) the bone--soft-tissue interface $\Gamma_{\mathrm{bnd}}$, where soft-tissue contacts the repositioned bony segments, and 
(ii) the facial outer surface $\Gamma_{\mathrm{face}}$, corresponding to the visible external skin surface.

To characterize surgical contact behavior, we partition the $\Gamma_{\mathrm{bnd}}$ nodes into three subsets: 
\emph{fixed} regions that remain fixed during surgery, 
\emph{moving} regions that are strictly following the prescribed rigid bone movement, and 
\emph{sliding} regions where non-rigid interface movement may occur. 
Let $V$ denote the set of all discretized nodes in $ \Omega_0$, which we partition into the interface node set $V_\mathrm{bnd}$ and the remaining free node $V_{\mathrm{free}} := V \setminus V_\mathrm{bnd}$, where mechanical equilibrium is enforced.
The input is a bone plan 
$\mathcal{P} = \{(R_k, t_k)\}_{k=1}^K$, 
representing the rigid transformations (rotation $R_k$, translation $t_k$) for $K$ bone segments. 
%
\begin{figure}[t]
    \centering
    \includegraphics[width=0.9\textwidth]{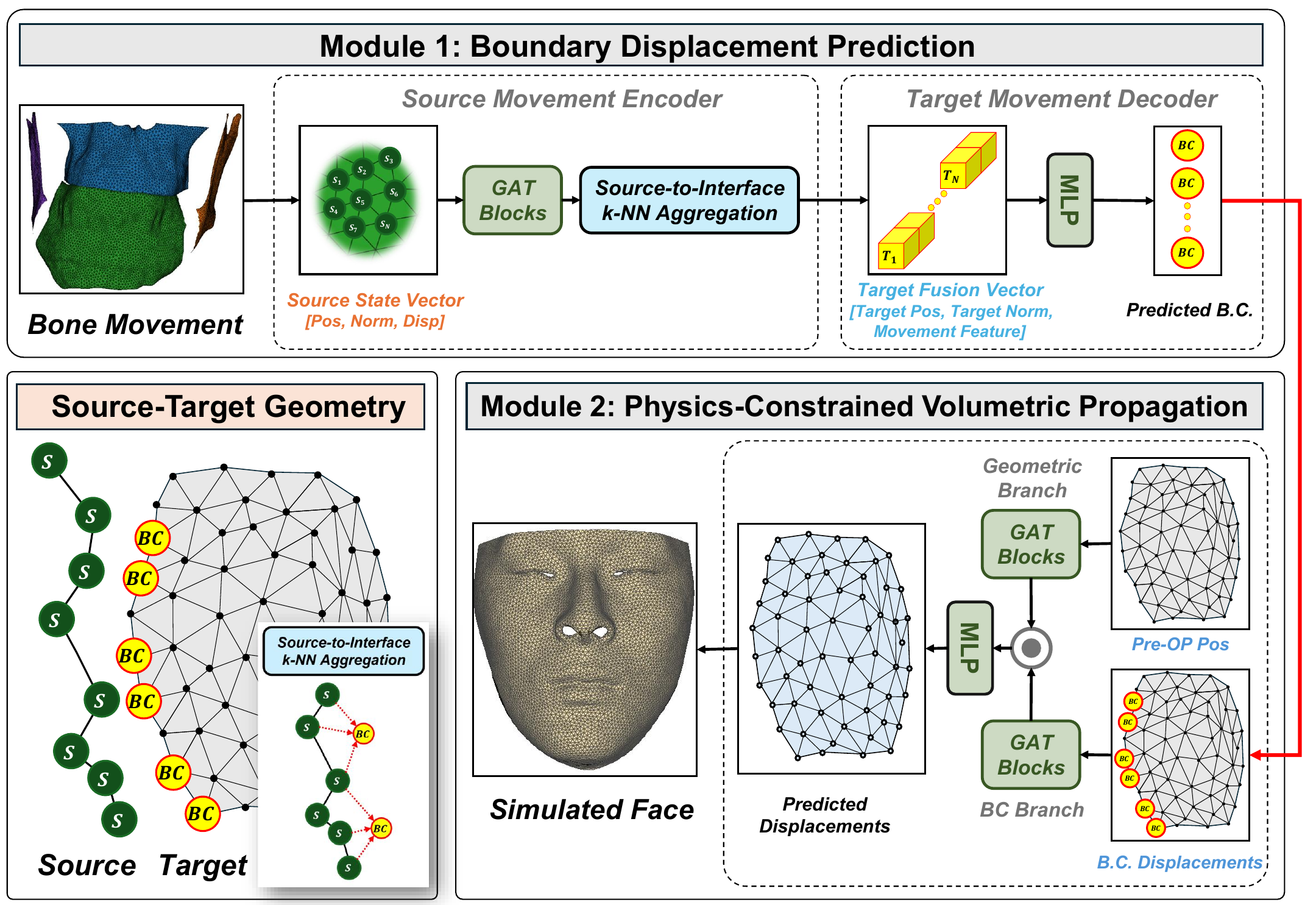}
    \caption{PINNOCHIO architecture. The boundary displacement prediction module estimates interface displacements from bone movements, and the volumetric propagation module propagates these displacements throughout the volume.}
    \label{fig:architecture}
\end{figure}

\section{PINNOCHIO Framework}
\subsection{Overview of the PINNOCHIO Framework}
To address the heterogeneous biomechanics of facial soft tissue, where discontinuous sliding at contact interfaces coexists with continuous hyperelastic deformation within the volume, PINNOCHIO adopts a hybrid sequential inference pipeline (\textbf{\Cref{fig:architecture}}). The framework consists of two coupled modules: \textbf{(1) Boundary Displacement Prediction}, which estimates non-rigid displacements on the bone--soft-tissue interface, and \textbf{(2) Physics-Constrained Volumetric Propagation}, which treats these predictions as boundary conditions to propagate deformation throughout the volume under physics-based constraints. To reflect heterogeneous tissue properties, we partition the preoperative volume $\Omega_0$ into two distinct anatomical layers: a deep muscular layer $\Omega_{\mathrm{mus}}$ (contractile tissues), and a superficial layer $\Omega_{\mathrm{sup}}$ (fat and skin).
\subsection{Network Modules}
\paragraph{\textbf{Boundary Displacement Prediction.}}
To predict boundary displacements $u_\mathrm{bnd}$ on the interface node set $V_\mathrm{bnd}$, we implement the predictor $P_\mathrm{bnd}$ as an encoder–decoder that maps a source-structure graph (bone or muscle) to interface-vertex displacements (\textbf{\Cref{fig:architecture}}). Because different interfaces are driven by different anatomical structures (e.g., bone–muscle, bone–superficial, and muscle–superficial), we adopt a flexible source–target design. First, a Source Encoder processes the relevant source structure using a Graph Attention Network (GAT)~\cite{brody2021attentive} to capture local geometry and varying kinematic patterns. Next, through Source-to-Interface k-NN Aggregation, we collect and apply max-pooling features from $k$ nearby source nodes to form a local context vector for each interface vertex. Finally, an MLP Decoder predicts the final boundary displacement by combining the aggregated features with a reference movement cue $c_\mathrm{ref}$. This cue embeds known surgical priors: zero for \emph{fixed} regions and rigid bone movement for \emph{moving} regions. Crucially, a binary reliability mask indicates where this cue is valid, guiding the network to focus entirely on learning the unknown, non-trivial \emph{sliding} regions.
\paragraph{\textbf{Volumetric Propagation.}}
The Volumetric Propagator $P_\mathrm{vol}$ propagates the predicted boundary displacement $u_\mathrm{bnd}$ throughout the soft-tissue domain to estimate a nodal displacement field $u$ over the entire volumetric mesh, including free nodes $V_\mathrm{free}$. We implement $P_\mathrm{vol}$ as a Dual-Branch GNN that fuses mesh geometry with boundary displacement cues. The Geometric Branch encodes the preoperative coordinates of the mesh, while the Boundary Condition (BC) Branch encodes the sparse boundary signal $u_\mathrm{sparse}$ (where $u_\mathrm{sparse}=u_\mathrm{bnd}$ on $V_\mathrm{bnd}$ and zero elsewhere) alongside a mask $\mathrm{m}$ indicating nodes with prescribed boundary displacements. Both branches utilize GAT blocks to aggregate local neighborhood information. The resulting embeddings are concatenated and passed through an MLP decoder to predict the final dense volumetric displacement field $\mathrm{u}$.
\subsection{Physics-Constrained Training of PINNOCHIO}
\paragraph{\textbf{Physics Loss.}}
To encourage biomechanical validity without internal ground truth, we model facial soft tissue as a Neo-Hookean hyperelastic material~\cite{simo1991quasi}. For each tetrahedral element with volume $V_e$, the deformation gradient is defined as $F_e = I + \nabla u$, where $\nabla u$ is the element-wise displacement gradient. The strain energy density is defined as:
\begin{equation}
\Psi(F_e) = \frac{\mu}{2}(\mathrm{tr}(F_e^\top F_e) - 3) 
- \mu \ln J_e 
+ \frac{\lambda}{2} (\ln J_e)^2,
\end{equation}
where $J_e = \det(F_e)$, and the Lamé parameters $\mu$ and $\lambda$ are spatially assigned to reflect heterogeneous tissue stiffness. We derive the first Piola–Kirchhoff stress $P_e = \frac{\partial \Psi}{\partial F_e}$~\cite{holzapfel2002nonlinear} to compute nodal internal forces $f_{\mathrm{int}}(v_i)$ using standard finite-element shape-function gradients. We enforce quasi-static equilibrium on free interior nodes $V_\mathrm{free}$ by penalizing the internal force residual:
\begin{equation}
\mathcal{L}_{\mathrm{phy}} =
\frac{1}{|V_{\mathrm{free}}|}
\sum_{v_i \in V_{\mathrm{free}}}
\| f_{\mathrm{int}}(v_i) \|_2^2.
\end{equation}
\paragraph{\textbf{Outer-Surface Loss.}}
Since dense point-to-point correspondence is unavailable for the clinically observed postoperative outer surface $\Gamma_{\mathrm{face}}$, we enforce surface agreement using the symmetric Chamfer Distance between the predicted surface ($S_\mathrm{pred}$) and post-operative surface ($S_\mathrm{post}$):
\begin{equation}
\mathcal{L}_{\mathrm{surf}} =
\mathcal{L}_{\mathrm{CD}}(S_\mathrm{pred}, S_\mathrm{post})
\end{equation}
\paragraph{\textbf{Boundary Consistency Loss.}}
Since the boundary displacement $u_\mathrm{bnd}$ is provided as a boundary displacement input to the volumetric propagator, we encourage the predicted displacement field $u$ to remain consistent with $u_\mathrm{int}$ on the boundary nodes $V_\mathrm{bnd}$, enforcing soft consistency rather than hard constraints:
\begin{equation}
\mathcal{L}_{\mathrm{bnd}} =
\frac{1}{|V_\mathrm{bnd}|}
\sum_{v_i \in V_\mathrm{bnd}}
\| u(v_i) - u_\mathrm{bnd}(v_i) \|_2^2.
\end{equation}
\paragraph{\textbf{Overall Training Objective.}}
The total loss balances these objectives with weighting coefficients ($\lambda_{\mathrm{phy}}$, $\lambda_{\mathrm{surf}}$, and $\lambda_{\mathrm{bnd}}$) that balance physical consistency, clinical surface agreement, and interface boundary consistency:
\begin{equation}
\mathcal{L}_{\mathrm{total}} =
\lambda_{\mathrm{phy}} \mathcal{L}_{\mathrm{phy}}
+
\lambda_{\mathrm{surf}} \mathcal{L}_{\mathrm{surf}}
+
\lambda_{\mathrm{bnd}} \mathcal{L}_{\mathrm{bnd}},
\end{equation}
\paragraph{\textbf{Physics-Enabled Sim-to-Real Adaptation.}}
To stabilize optimization, we first pretrain PINNOCHIO on FEM-simulated data, replacing the outer-surface loss with a volumetric mean-squared error (MSE) against ground-truth displacements. We then fine-tune end-to-end on real postoperative cases using the overall objective ($\mathcal{L}_{\mathrm{total}}$), relying on the outer-surface loss ($\mathcal{L}_{\mathrm{surf}}$) alongside the physics and boundary constraints.
\section{Experiments \& Results}
\subsection{Experimental Setup}
We evaluated PINNOCHIO on 40 clinical orthognathic surgery cases using five-fold cross-validation. Each case includes a planned rigid movement of four bone segments (LeFort I, mandibular distal, and bilateral distal segments). From pre-operative CT scans, we obtained anatomical segmentations (bone, muscle, and superficial tissue) using Skull-Engine~\cite{liu2021skullengine}, and further refined to reflect the surgical movements. The bone was represented as a triangular surface mesh, while muscle and superficial tissue were tetrahedral volumetric meshes. Post-operative outer facial surfaces were extracted from 3dMD scans for supervision.

We compared PINNOCHIO against a high-fidelity FEM simulator (FEM-RLSE~\cite{kim2021novel}) and representative deep learning baselines (ACMT-Net~\cite{fang2024correspondence}, DGCFP~\cite{huang2025maxillofacial}). Surface accuracy was evaluated using Chamfer Distance (CD) and Hausdorff Distance (HD), and Patient-level Within-threshold Rate (PWR, \%), defined as the percentage of the facial surface area where the prediction error falls below a predefined clinical threshold (e.g., 2 mm), averaged across all patients. Mesh validity was assessed using the minimum Jacobian determinant ($J$) for element inversion, and Equilibrium Residual (EqRes) for physical consistency. Statistical significance was evaluated using the Wilcoxon signed-rank test~\cite{wilcoxon1992individual}. Loss weights $(\lambda_{\mathrm{phy}}, \lambda_{\mathrm{surf}}, \lambda_{\mathrm{bnd}})$ were empirically set to $(1.0, 2.0, 1.0)$. Hyperelastic parameters were set per tissue type, with Young's modulus $E = 6$ kPa  and Poisson's ratio $\nu=0.49$ for muscle, and $E = 4$ kPa, $\nu=0.49$ for the superficial layer. Inference time was measured on a NVIDIA DGX-1 server with V100 GPUs. 

\begin{table}[t]
\centering
\caption{Quantitative comparison on whole-face surface accuracy and inference time. $\uparrow$/$\downarrow$: higher/lower is better. Best results are in bold.}
\label{tab:main_results}
\scriptsize
\setlength{\tabcolsep}{2.5pt}
\renewcommand{\arraystretch}{1.00}

\begin{tabularx}{\linewidth}{
l
>{\centering\arraybackslash}p{0.62in}
>{\centering\arraybackslash}p{0.62in}
>{\centering\arraybackslash}X
>{\centering\arraybackslash}X
>{\centering\arraybackslash}p{0.55in}}
\toprule
Method & CD(mm)$\downarrow$ & HD(mm)$\downarrow$ & PWR@1mm(\%)$\uparrow$ & PWR@2mm(\%)$\uparrow$ & Time(s)$\downarrow$ \\
\midrule
ACMT-Net~\cite{fang2024correspondence}      & $2.19\!\pm\!0.79$\textbf{*} & $5.56\!\pm\!2.35$\textbf{*} & $36.04\!\pm\!12.98$\textbf{*} & $63.59\!\pm\!14.97$\textbf{*} & $\mathbf{0.10}$ \\
DGCFP~\cite{huang2025maxillofacial}     & $1.71\!\pm\!0.62$\textbf{*} & $4.94\!\pm\!2.03$\textbf{*} & $47.28\!\pm\!12.36$\textbf{*} & $72.43\!\pm\!12.18$\textbf{*} & 0.54 \\
FEM-RLSE~\cite{kim2021novel}  & $1.30\!\pm\!0.29$\textbf{*} & $3.16\!\pm\!0.80$\textbf{*} & $50.83\!\pm\!10.97$\textbf{*} & $80.90\!\pm\!10.12$\textbf{*} & $1.26{\times}10^{4}$ \\
\textbf{PINNOCHIO} & $\mathbf{1.12\!\pm\!0.26}$ & $\mathbf{2.73\!\pm\!0.69}$ & $\mathbf{58.36\!\pm\!12.60}$ & $\mathbf{86.55\!\pm\!8.42}$ & $3.24$ \\
\bottomrule
\end{tabularx}
\end{table}

\newcolumntype{Y}{>{\centering\arraybackslash}X}

\begin{table}[t]
\centering
\caption{Incremental ablation of decomposition and physics regularization. Lower is better for CD and EqRes, while higher is better for PWR and J.}
\label{tab:ablation_incremental}
\scriptsize
\setlength{\tabcolsep}{2.5pt}
\renewcommand{\arraystretch}{1.00}

\begin{tabularx}{\linewidth}{
>{\centering\arraybackslash}p{0.7in}
>{\centering\arraybackslash}p{0.65in}
>{\centering\arraybackslash}X
>{\centering\arraybackslash}X
>{\centering\arraybackslash}X
>{\centering\arraybackslash}X}
\toprule
 & & \multicolumn{2}{c}{Surface Distance} & \multicolumn{2}{c}{Mesh Validity} \\
\cmidrule(lr){3-4}\cmidrule(lr){5-6}
\raisebox{1.5ex}[0pt]{Decomposition} & \raisebox{1.5ex}[0pt]{Physics Loss} & CD(mm)$\downarrow$ & PWR@1mm(\%)$\uparrow$ & EqRes$\downarrow$ & J$\uparrow$ \\
\midrule
 & \checkmark & $1.04\!\pm\!0.27$\textbf{*} & $62.35\!\pm\!11.74$\textbf{*} & $0.25\!\pm\!0.34$ & $0.86\!\pm\!0.04$ \\
\checkmark & & $1.02\!\pm\!0.24$\textbf{*} & $64.10\!\pm\!12.63$ & $1.73\!\pm\!1.14$\textbf{*} & $0.68\!\pm\!0.14$\textbf{*} \\
\checkmark & \checkmark & $\mathbf{0.98\!\pm\!0.22}$ & $\mathbf{64.88\!\pm\!11.58}$ & $\mathbf{0.20\!\pm\!0.34}$ & $\mathbf{0.87\!\pm\!0.05}$ \\
\bottomrule
\end{tabularx}
\end{table}

\subsection{Results}
\begin{figure}[t]
    \centering
    \includegraphics[width=0.84\textwidth]{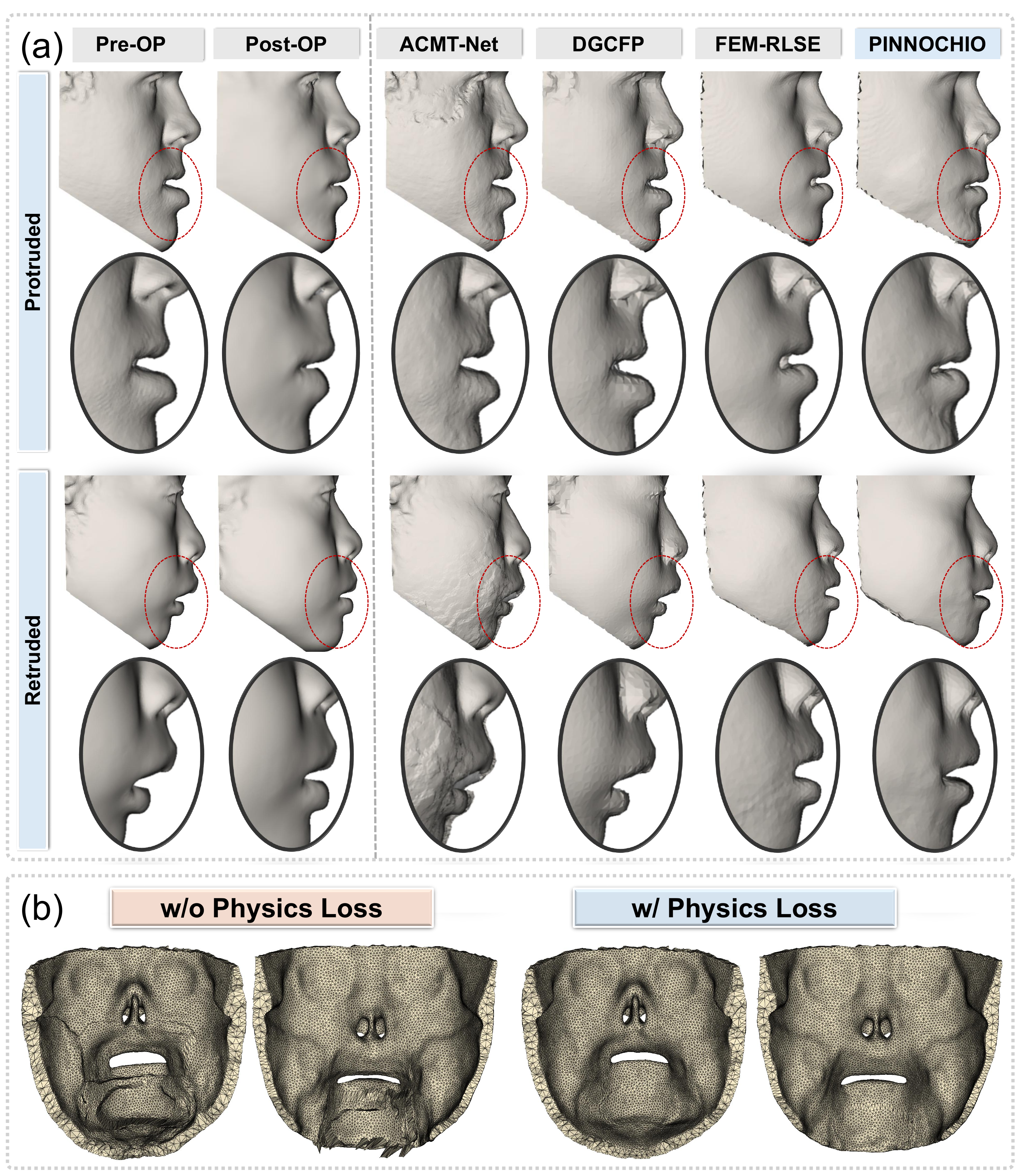}
    \caption{Qualitative results. \textbf{(a)} Predicted results compared with baselines and ground truth on representative cases. \textbf{(b)} Impact of physics-based regularization ($\mathcal{L}_{\mathrm{phy}}$). Posterior views show how physics constraints suppress element inversion and artifacts.}
    \label{fig:qualitative_results}
\end{figure}

\textbf{\Cref{tab:main_results}} compares PINNOCHIO with FEM (FEM-RLSE) and deep learning baselines (ACMT-Net, DGCFP). Our method achieved the lowest surface error (CD: $1.12$ mm, HD: $2.73$ mm) and highest PWR@2mm ($86.55$\%). Furthermore, PINNOCHIO required only $3.24$ seconds per case, compared to $3.5$ hours ($1.26\times 10^4$ s) for FEM-RLSE, achieving a substantial speedup while maintaining superior accuracy. Qualitatively (\textbf{\Cref{fig:qualitative_results}(a)}), PINNOCHIO's predictions closely resemble the post-operative ground truth on par with FEM-RLSE across both protruded and retruded patient profiles, accurately capturing complex soft-tissue deformations where deep learning baselines fail.

\textbf{\Cref{tab:ablation_incremental}} ablates the sequential decomposition against a unified PINN structure that jointly models the interface and volumetric domains. This decomposition yielded meaningful improvements in surface accuracy and physical equilibrium (EqRes), demonstrating that decoupling heterogeneous mechanics facilitates more effective learning. Additionally, removing the physics loss increased surface errors and degraded volumetric mesh validity ($J$), indicating reduced biomechanical consistency. As shown in \textbf{\Cref{fig:qualitative_results}(b)}, relying solely on outer-surface supervision produced geometrically unstable internal deformations, whereas the physics term ensured smooth, physically coherent volumes.

\section{Conclusions}
We presented PINNOCHIO, a physics-informed framework for facial soft-tissue simulation in orthognathic surgery. By sequentially decomposing interface prediction and volumetric propagation, our approach enables stable learning under surface-only supervision. Achieving competitive accuracy and a substantial speedup over FEM, it facilitates real-time iterative planning. Future work will incorporate broader surgical variations (e.g., genioplasty) and integrating patient-specific material estimation (e.g., via ultrasound) to further enhance biomechanical fidelity. Overall, PINNOCHIO provides a practical and extensible basis for clinically deployable surgical planning.
\subsubsection{Acknowledgments.}
This work was supported by the National Institutes of Health (NIH) under awards R01DE027251 and R01DE021863.

\subsubsection{Disclosure of Interests.}
No competing interests declared.
%
%
%
%
\bibliographystyle{splncs04}
\bibliography{mybibliography}
\end{document}